\documentclass[twocolumn,superscriptaddress,nofootinbib,aps,prl,floatfix,preprintnumbers,amsmath,amssymb,groupedaddress]{revtex4}

\usepackage{dsfont}
\usepackage{epsfig}
\usepackage{slashed}
\usepackage{bbold}
\usepackage{psfrag}
\usepackage{color}
\PassOptionsToPackage{caption=false}{subfig}
\usepackage{subfig}
\usepackage{multirow}
\usepackage{booktabs}
\usepackage{hyperref}

\newcommand{\be}{\begin{equation}}
\newcommand{\ee}{\end{equation}}
\newcommand{\bea}{\begin{eqnarray}}
\newcommand{\eea}{\end{eqnarray}}

\newcommand{\TeV}{\textrm{ TeV}}

\begin{document}
%\title{TeV SUSY from Gravitino Dark Matter}
\title{A Cosmological Upper Bound on Superpartner Masses}

\author{Lawrence J. Hall}
\affiliation{Department of Physics, University of California, Berkeley, CA 94720, USA}
\affiliation{Theoretical Physics Group, Lawrence Berkeley National Laboratory, Berkeley, CA 94720, USA}
\author{Joshua T. Ruderman}
\affiliation{Department of Physics, University of California, Berkeley, CA 94720, USA}
\affiliation{Theoretical Physics Group, Lawrence Berkeley National Laboratory, Berkeley, CA 94720, USA}
\author{Tomer Volansky}
\affiliation{Raymond and Beverly Sackler School of Physics and Astronomy, Tel-Aviv University, Tel-Aviv 69978, Israel}

%\pacs{PACS}
%\keywords{keywords}
%\preprint{UCB-PTH-13/?? }
\begin{abstract}

If some superpartners were in thermal equilibrium in the early universe, and if the lightest superpartner is a cosmologically stable gravitino, then there is a powerful upper bound on the scale of the superpartner masses.   
Typically the bound is below tens of TeV, often much lower, 
and has similar parametrics to the WIMP miracle. 
\end{abstract}
\maketitle
%%%%%%%%%%%%%%%%%%%%%%%%%%%%%%%%%%%%%%%%%%%%%%

%%%%%%%%%%%%%%%%%%%%%%%%%%%
{\noindent\bf INTRODUCTION.} 
%%%%%%%%%%%%%%%%%%%%%%%%%%%
A natural weak scale, precision gauge coupling unification and dark matter provide powerful arguments for weak-scale supersymmetry.    However, to date, direct evidence for supersymmetry is still missing and thus whether or not low-scale supersymmetry is realized in nature remains unknown.   In fact, the recent discovery of a 125 GeV Higgs boson~\cite{:2012gk,:2012gu} implies a fine-tuning in the MSSM worse than 1\%~\cite{Hall:2011aa}, and   searches for supersymmetry at the LHC are placing limits on colored superpartners in the region of 1 TeV~\cite{:2013wc, :2012rz}.    Therefore, the superpartner mass scale, $\tilde m$, may be decoupled from the weak scale and could, in principle, be anywhere  between the present experimental bound near 1 TeV up to the Planck scale.  With the naturalness reasoning aside, the question arises: {\em Are there arguments for superpartners at the TeV scale that are unrelated to the stabilization of the weak scale?}

The argument for TeV superpartners from gauge coupling unification alone is weak, as logarithmic running implies that the precision changes only mildly as $\tilde{m}$ increases well above 1 TeV.  On the other hand, there is a powerful and well-known argument for TeV-scale superpartners from the cosmological abundance of the lightest supersymmetric particle (LSP)~\cite{Goldberg:1983nd}.
 This results from LSP freeze-out and follows from three assumptions:
\begin{enumerate}
\item[(i).] The LSP is cosmologically stable. 
\item[(ii).] The reheat temperature of visible particles after inflation, $T_R$,  was sufficiently high, $T_R \gtrsim \tilde m$.
\item[(iii).] There is no substantial late-time dilution of the LSP abundance.
\end{enumerate}
The second assumption implies that the standard model superpartners were in thermal equilibrium.  If we further assume
\begin{enumerate}
\item[(iv-A).] The LSP reached thermal equilibrium,
\end{enumerate}
then the thermal freeze-out relic abundance leads to the overclosure bound on the LSP mass,
\be
m_{LSP}^2 \, \leq \, 2.0 \, \frac{23}{x_f} \; \alpha_{\rm eff}^2  \; T_{\rm eq} M_{\rm Pl} \, \simeq \, \left( 2.3 \TeV  \; \frac{\alpha_{\rm eff}}{0.03} \right)^2\,,
\label{eq:mlspfromFO}
\ee
where $T_{\rm eq} \simeq 1.5$~eV is the temperature of matter-radiation equality, $M_{\rm Pl}\simeq 2.4\times10^{18}$~GeV is the reduced Planck mass, and $x_f$ is $\tilde{m}$ divided by the freeze-out temperature.   The coupling strength $\alpha_{\rm eff}$, appearing in the thermally averaged  LSP annihilation cross section at freeze-out, is defined by $\langle\sigma v\rangle = 4 \pi \alpha_{\rm eff}^2/\tilde{m}^2$. It is 0.03 (0.01) for wino (Higgsino) LSP, but can have a much larger variation. 

The TeV scale from freeze-out thus results parametrically as the geometric mean of $T_{\rm eq}$ and $M_{Pl}$ and is independent of the weak scale.
The equality holds for LSP dark matter.
Eq.~\eqref{eq:mlspfromFO} is a very important result: in models where superpartner masses are characterized by a single scale, $\tilde{m}$ is likely in the 1-10 TeV window, and in Split Supersymmetry~\cite{ArkaniHamed:2004fb} the fermonic superpartners lie in the TeV region.

\begin{figure*}[t!]
\centering
\includegraphics[width=0.7\linewidth]{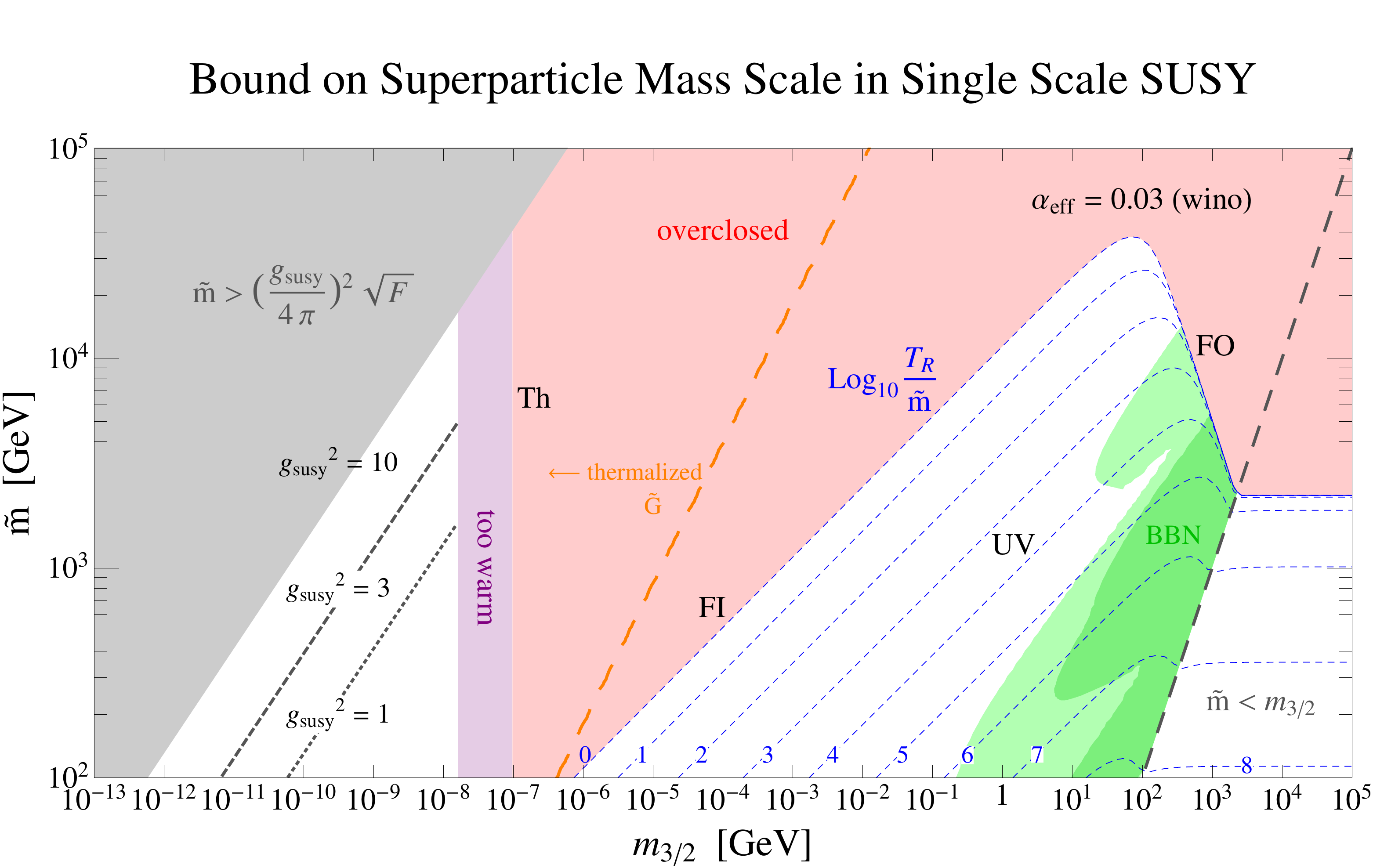}
\caption{\label{fig:SingleScale} The cosmologically allowed region in the $(m_{3/2},\tilde m)$ plane, for a single scale SUSY with an LSP adhering  to assumptions (i)-(iv-B) discussed in the text.  The gravitinos are (are not) thermalized to the left (right) of the {\bf orange} dashed line (assuming $T_R = \tilde m$).  Even when $T_R$ is as low as $\tilde{m}$, gravitinos provide too much dark matter in the {\bf red} region, which has borders labelled by the relevant process Th, FI or FO.   As $T_R$ is increased the overclosed region becomes larger, as illustrated by the dashed {\bf blue }lines, because UV scattering at $T_R$ produces more gravitinos than freeze-in.  At the edge of the red region (suitably enlarged for $T_R > \tilde{m}$) gravitinos provide the observed dark matter.
 In the region to the right of the slanted {\bf black} dashed line the gravitino is not the LSP; this is the conventional WIMP LSP freeze-out region, with a limit of 2.3 TeV for a wino LSP.  The {\bf green} region is excluded by the effects of late decays of LOSPs to gravitinos during big bang nucleosynthesis (BBN)~\cite{Jedamzik:2006xz}; {\bf light green} shading corresponds to a neutral LOSP with 100\% hadronic BR, and {\bf dark green} shading to a neutral LOSP with 1\% hadronic BR and 99\% electromagnetic BR.   The BBN limits when the gravitino is not the LSP are model dependent  and are not shown~\cite{Kawasaki:2008qe}.  The {\bf purple} region next to the ``Th" contour is excluded as the gravitino component of dark matter is too warm \cite{Viel:2005qj}.   
The {\bf gray} shading (and corresponding gray dashed and dotted lines) shows the regions with $g_{susy}^2 > 10 \; (3,1)$, which are excluded as described in the text.   }
\end{figure*}

From the above list of assumptions it is clear how to evade the bound in Eq.~\eqref{eq:mlspfromFO}, allowing $\tilde{m}$ many orders of magnitude above the TeV scale.  In particular, violating assumption (i) through, e.g., R parity breaking, may void the bound entirely.   In that case, DM could arise, for instance, from a hidden sector or from axions.    Violating assumption (ii), having $\tilde{m}$ well above $T_R$, allows the superpartners to have no cosmological role, hence evading the bound.   Finally, assumption (iii) may not hold if additional late-decaying states reheated the universe.

In this paper we study the intriguing possibility of violating assumption (iv-A).  
 Indeed, there are numerous scenarios where DM  is only very weakly coupled so that its abundance does not follow from thermal freeze-out, invalidating Eq.~\eqref{eq:mlspfromFO}.

The most common scenario of this kind has (iv-A) replaced by
\begin{enumerate}
\item[(iv-B).]  The gravitino is the LSP (and the Lightest Observable-sector SuperPartner (LOSP) decays predominantly to gravitinos). 
\end{enumerate}
The gravitino, present in all supersymmetric theories, has interactions that are highly constrained and very weak.
The gravitino has a cosmological abundance determined by thermal scattering, freeze-in, and freeze-out and decay,  and reaches thermal equilibrium, in accordance with (iv-A), only when it is very light.   The gravitino abundance has been studied in detail for the case of weak-scale superpartners, for example leading to bounds on $T_R$ as a function of the gravitino mass \cite{Moroi:1993mb}.   In this letter, however, we take a different approach and {\em derive the cosmological bound on the superpartner mass scale for a gravitino LSP. }  We find this bound to be strong, so that under the quite mild assumptions of (i), (ii), (iii) and (iv-A) {\it or} (iv-B), supersymmetry, if it exists, must be in the (multi-) TeV domain.   We also derive bounds for the split spectrum case and scenarios where the LOSP does not predominantly decay to the gravitino.

%%%%%%%%%%%%%%%%%%%%%%%%%%%%%%%%%%%%%%
{\noindent\bf SINGLE SCALE SUSY}.
%%%%%%%%%%%%%%%%%%%%%%%%%%%%%%%%%%%%%%
In this section we take all superpartners of the observable sector to be characterized by a single mass scale, $\tilde{m}$, and leave the case of a non-degenerate spectrum to the next sections.   Our aim is to derive a general bound on the scale $\tilde{m}$ from overproduction of  gravitinos.  We ignore other possible components to DM since they would only strengthen the bound.   A key superpartner is the LOSP, 
since it undergoes freeze-out.  We allow a very wide variation in the $(m_{3/2}, \tilde{m}, T_R)$ space.  

The upper bound on $\tilde{m}$ follows from the three assumptions (i), (ii) and (iii).  
Assumption (ii) implies that the observable sector produces gravitinos from three sources: gaugino scattering at $T_R$~\cite{Bolz:2000fu, Pradler:2006qh, Rychkov:2007uq}, $Y_{3/2}^{UV}$,  gravitino ``freeze-in" from decays of visible sector superpartners at $T \sim \tilde{m}$~\cite{Hall:2009bx, Cheung:2011nn}, $Y_{3/2}^{FI}$, and LOSP freeze-out and decay~\cite{Feng:2003uy}, $Y_{3/2}^{FO}$.  For sufficiently small $m_{3/2}$, the gravitinos are in thermal equilibrium when $T=\tilde{m}$; in this case $Y_{3/2}^{UV}+Y_{3/2}^{FI}$ are replaced by a thermal abundance, and $Y_{3/2}^{FO}$ may be neglected.  
Below, in accordance with assumption (iv-B), we assume the LOSP branching ratio to the gravitino is ${\cal O}(1)$.   In the final section we discuss how our bound is weakened when this assumption is relaxed. Gravitinos may also be produced from other sectors or they may arise from an initial condition~\cite{Kallosh:1999jj, Giudice:1999yt}.  However, these additional sources of gravitinos only strengthen our bound, and to be conservative we ignore them. 

\begin{figure*}[t!]
\centering
\includegraphics[width=0.8\linewidth]{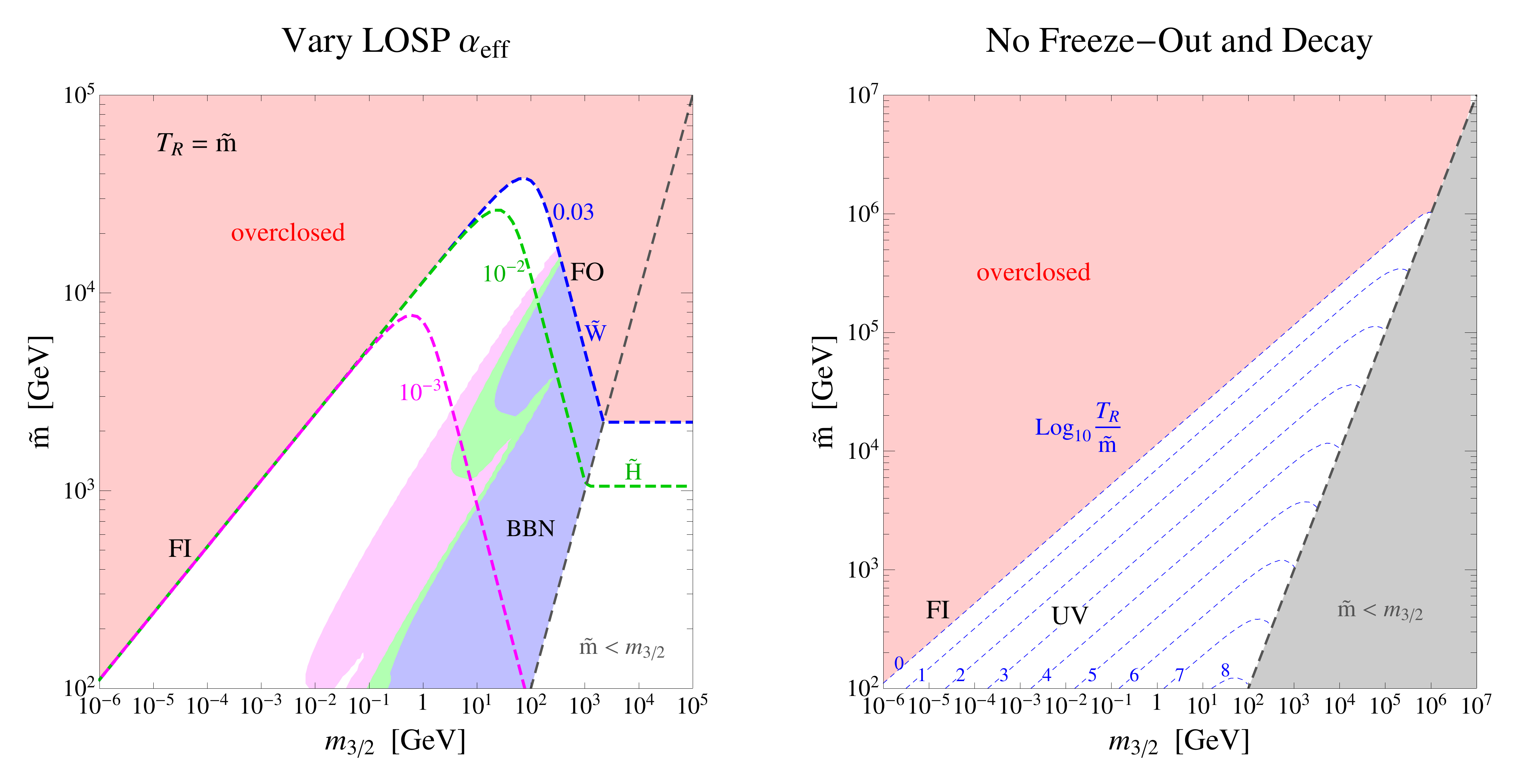}
\caption{\label{fig:SingleScaleVariations}   
{\bf Left}: The bound on $\tilde{m}$ in the single-scale SUSY case, for $\alpha_{\rm eff} = 0.03, 10^{-2}$ and $10^{-3}$ in blue, green and purple respectively, assuming $T_R = \tilde{m}$.  As $\alpha_{\rm eff}$ decreases freeze-out yields a larger abundance, so the FO boundary and the BBN constraints (shown shaded in the corresponding colors) both become more stringent.  As $T_R$ is raised, the bounds become more stringent as indicated by the blue dashed lines of Fig.~\ref{fig:SingleScale}.  {\bf Right}:  The bound on $\tilde{m}$ when the contribution to the gravitino abundance from freeze-out and decay is negligible.   This may be the case in several scenarios, as discussed in the last section.   The dashed blue lines demonstrate the strengthening of the bound as $T_R$ is increased.  We do not analyze the region with $\tilde{m} < m_{3/2}$ as the results are model-dependent.}
\end{figure*}

If gravitinos do not thermalize, the condition that they not yield too large a DM abundance is
\be
C_{UV} \, \frac{T_R \tilde{m}^2}{m_{3/2}} + C_{FI} \, \frac{\tilde{m}^3}{m_{3/2}} +  C_{FO} \, \frac{\tilde{m} m_{3/2}}{\alpha_{\rm eff}^2} \; \leq \; a M_{\rm Pl} T_{\rm eq} \,,
\label{eq:gravDM}
\ee
where $a=0.27$ and $\alpha_{\rm eff}$ is now the coupling relevant for LOSP annihilation.  The three terms labelled UV, FI and FO correspond to scattering at $T_R$, freeze-in and freeze-out and decay and occur with rate constants $C_{UV} = \gamma_3  \, \frac{15 \sqrt{90}}{2 \pi^3 {g_*}^{3/2}} \simeq 2.4 \times 10^{-4}$,  $C_{FI} = \frac{405}{2 \pi^4}\sqrt{\frac{5}{2}} \frac{1}{g_*^{3/2}} \frac{n_{FI}}{4 \pi} \simeq 3.8 \times 10^{-4}$ and $C_{FO} =  \frac{3 \sqrt{5} \, x_f}{8  \sqrt {2 \, g_*} \pi^2} \simeq 0.13 \left( \frac{x_f}{23} \right)$. 
Here $\gamma_3\simeq 0.36$ is related to the thermal corrections of the scattering process~\cite{Rychkov:2007uq}, $g_* = 228.75$, and $n_{FI}$ counts the number of fermions and complex scalars participating in the freeze-in with mass $\tilde m$;  with degenerate MSSM sparticles, $n_{FI} = 36+9+12+4=61$.  The equality in Eq.~(\ref{eq:gravDM}) corresponds to the case that these processes yield the observed DM abundance.  
If gravitinos do thermalize, the overabundance constraint becomes~\cite{Pagels:1981ke}
\be
C_{Th} \, m_{3/2}  \; \leq \; a \, T_{\rm eq}\,,
\label{eq:gravDMth}
\ee
with $C_{Th} = Y_{\gamma} = 45\xi(3)/\pi^4 g_{*s}\approx 2.4\times 10^{-3}$.  Here $g_{*s} \simeq g_* = 228.75$.   
The resulting bound on $\tilde{m}$ as a function of $m_{3/2}$ is shown in Fig.~\ref{fig:SingleScale} for $\alpha_{\rm eff} = 0.03$, relevant for a (perturbative) wino LOSP.   We do not include the non-perturbative Sommerfeld effect~\cite{Hisano:2006nn}, which results in an $\mathcal{O}(1)$ shift in $\alpha_{\rm eff}$.  

 When gravitinos are not thermalized, the key point is the differing dependences of the three terms in Eq.~\eqref{eq:gravDM} on $\tilde{m}$ and $m_{3/2}$.  While all three terms have a positive power of $\tilde{m}$, the UV and FI terms are proportional to $1/m_{3/2}$ while the FO term is proportional to $m_{3/2}$, leading to contours in  Fig.~\ref{fig:SingleScale} with slopes of opposite signs.  Hence there is an upper bound,
\be
\tilde m^2 \; \leq \;
\frac{a/2}{\sqrt{C_{FO} C_D}}
 \,\, \alpha_{\rm eff} \, M_{Pl} \, T_{eq}\,,
\label{eq:mtildebound}
\ee
where $C_D =C_{UV}(T_R / \tilde m) + C_{FI}$. 
At the bound $m_{3/2}=\sqrt{C_D/C_{FO}} \, \alpha_{\rm eff} \, \tilde{m}$. For $T_R \gg \tilde{m}$ the bound becomes $ \tilde{m} \leq 27 \, \mbox{TeV} \, [(T_R / \tilde m)/10]^{-1/4}$ for $\alpha_{\rm eff} =0.03$ which weakens to $\tilde m \lesssim 38~\mathrm{TeV}$  for $T_R = \tilde{m}$.
Decreasing $\alpha_{\rm eff}$ makes the FO term larger, as shown in the left panel of Fig.~\ref{fig:SingleScaleVariations}  for $T_R = \tilde{m}$.  The parametrics of Eq.~(\ref{eq:mtildebound}) is similar, but not identical, to that in the so-called ``WIMP Miracle", Eq.~(\ref{eq:mlspfromFO}). 

A second allowed region occurs at very low $m_{3/2}$ in Fig.~\ref{fig:SingleScale}, where the gravitinos are thermalized for any $T_R 
\geq \tilde{m}$.   Here the bound on $\tilde{m}$ arises from theory rather than cosmology: $\tilde{m} \leq (g_{susy}/4 \pi)^2 \sqrt{F}$, where $g_{susy}$ is the strength of the coupling between obervable and supersymmetry breaking sectors, and $F= \sqrt 3 m_{3/2} M_{\rm Pl}$ is the supersymmetry breaking scale.  The bound results when the messenger scale takes its minimal value of $\sqrt{F}$, and is shown in Fig.~\ref{fig:SingleScale} for  $g_{susy}^2 = 1, 3$ and 10.  We note that it may be possible to construct realistic models of composite quarks and leptons having non-perturbative couplings, $g_{susy} \sim 4 \pi$ \cite{ArkaniHamed:1997fq}.

%%%%%%%%%%%%%%%%%%%%%%%%%%%%%%%%%%%%%%
{\noindent\bf NON-DEGENERATE SPECTRUM}. 
%%%%%%%%%%%%%%%%%%%%%%%%%%%%%%%%%%%%%%
\begin{figure*}[t!]
\centering
\includegraphics[width=0.99\linewidth]{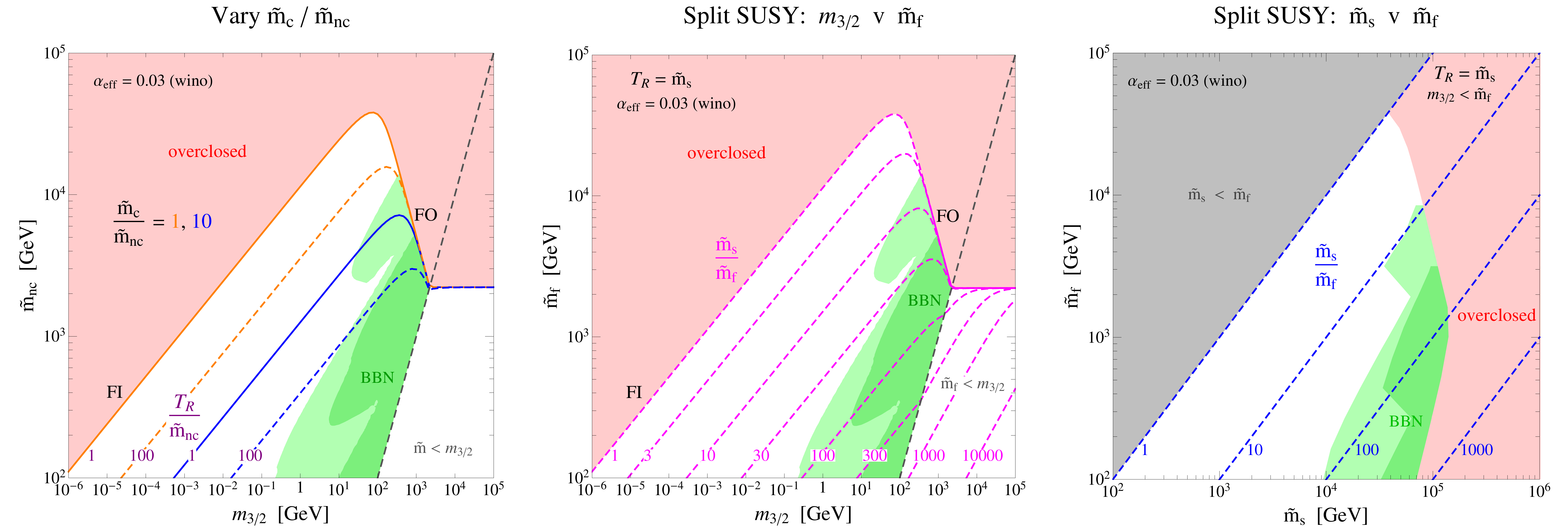}
\caption{\label{fig:Split} 
{\bf Left}: Bounds in the $(m_{3/2},\tilde m_{nc})$ plane for colored (non-colored) superpartners with mass $\tilde m_c$ ($\tilde m_{nc}$).  The importance of freeze-in as $m_c/m_{nc}$ is raised from 1 to 10 is seen by comparing the orange and blue lines.  The solid and dashed lines show the effect of increasing $T_R$ by 100.   {\bf Center}:  Similar to the left panel, the changes to the bound of Fig.~\ref{fig:SingleScale} is shown for the case of split-SUSY, where the scalar superpartner masses, $\tilde{m}_s$, are raised above the fermionic superpartner masses, $\tilde{m}_f$.   {\bf Right}: The overclosure bound  in the ($\tilde{m}_s,\tilde{m}_f)$ plane is shown for the split-SUSY case, where the gravitino mass has been chosen at each point to maximize the allowed region.  For split-SUSY $T_R = \tilde{m}_s$. In all panels the green shading is as in Fig.~\ref{fig:SingleScale}.  }  
\end{figure*}
The completely degenerate spectrum discussed above is special because non-degeneracies typically arise from renormalization group effects or the dynamics of the mediation of supersymmetry breaking.   How do non-degeneracies affect the above bounds?    

Non-degeneracies induce independent changes in the three gravitino production mechanisms.  The freeze-in process is dominated by the heaviest superpartners, $\tilde{m}_+$, and is suppressed compared to the degenerate case by $n_{FI}^+/n_{FI}$, where $n_{FI}^+$ is the number of these heavy superpartners.  The scattering process, 
dominated by gluino scattering, is proportional to the square of the gluino mass, $M_3^2$.  Finally, the freeze-out abundance is proportional to the LOSP mass, $\tilde{m}_-$, with $\langle\sigma v\rangle = 4 \pi \alpha_{\rm eff}^2/\tilde{m}_-^2$,  so that Eq.~\eqref{eq:gravDM} becomes  
\be
C_{UV} \frac{T_R M_3^2}{m_{3/2}} + \frac{C_{FI} n_{FI}^+}{n_{FI}} \frac{\tilde{m}_+^3}{m_{3/2}} +  C_{FO} \frac{\tilde{m}_- m_{3/2}}{\alpha_{\rm eff}^2}  \leq  a M_{\rm Pl} T_{\rm eq}.
\label{eq:gravDMnondegen}
\ee
While pure FO of Eq.~(\ref{eq:mlspfromFO}) bounds $m_{LSP}$, with a gravitino LSP the bound depends on the $m_{LOSP}, M_3$, and the mass dominating FI.

As a simple example, on the left of Fig.~\ref{fig:Split} we show the bound that results by taking all colored states at $\tilde{m}_c = \tilde{m}_+$ and all non-colored states at $\tilde{m}_{nc} = \tilde{m}_-$, assuming all superpartners are reheated.   As can be seen, the bound on $\tilde{m}_{nc}$  becomes much more stringent as $\tilde{m}_c$ is raised, being reduced to 7 TeV for $\tilde{m}_c/\tilde{m}_{nc} = 10$.    
Much of the allowed regions in Figs.~\ref{fig:SingleScale}, \ref{fig:SingleScaleVariations}-Left and \ref{fig:Split}-Left are within the LHC reach.

%%%%%%%%%%%%%%%%%%%%%%%%%%%%%%%%%%%%%%
{\noindent\bf SPLIT SUSY}. 
%%%%%%%%%%%%%%%%%%%%%%%%%%%%%%%%%%%%%%
In the split-SUSY  scenario~\cite{ArkaniHamed:2004fb}, 
where the scalar superpartner mass, $\tilde{m}_s$, becomes much larger than the fermionic superpartner mass, $\tilde{m}_f$, a bound on $\tilde{m}_f$, with a gravitino LSP, was discussed in~\cite{ArkaniHamed:2004yi}.  The freeze-in process dominates over the scattering process as long as 
$T_R > \tilde{m}_s$ \cite{ArkaniHamed:2004yi,Hall:2012zp}.   Using Eq.~\eqref{eq:gravDMnondegen}, with $\tilde{m}_s = \tilde{m}_+$ and $\tilde{m}_f = \tilde{m}_-$, yields the bound on $\tilde{m}_f$ shown in the center panel of Fig.~\ref{fig:Split} for various values of $\tilde{m}_s / \tilde{m}_f$.   To compute the bound, the split-SUSY 1-loop RGEs were used~\cite{Arvanitaki:2004eu, Giudice:2004tc}.
The bound on $\tilde{m}_s$ is in the region of 100 TeV, as shown in the right panel of Fig.~\ref{fig:Split}, and hence arbitrary flavor and CP violation in the squark mass matrix requires $T_R < \tilde{m}_s$.  Finally, we note that if $T_R$ is indeed below  $\tilde{m}_s$ a bound on $\tilde{m}_f$ may still be obtained, and is similar to that shown in Fig.~\ref{fig:SingleScale} up to ${\cal O}(1)$ corrections stemming from the absence of some diagrams in the finite-temperature thermal production of the gravitinos~\cite{Bolz:2000fu}.

The non-degeneracies explored in the left and center panels of Fig.~\ref{fig:Split} lead to similar bounds, and forbid large splittings between the light and heavy states (assuming that both are reheated). Indeed, as the splittings  increase, the BBN bounds rapidly become very constraining.

%%%%%%%%%%%%%%%%%%%%%%%%%%%%%%%%%%%%%%
{\noindent\bf RELAXING ASSUMPTION (iv-B)}. 
%%%%%%%%%%%%%%%%%%%%%%%%%%%%%%%%%%%%%%
We now consider how the bound on superparticle masses is relaxed in theories that violate assumption (iv-B).

LOSP freeze-out and decay may not produce a significant yield of LSP gravitinos, depleting $Y_{3/2}^{FO}$.  This occurs, for example, if the LOSP dominantly decays through $R$-parity violating (RPV) operators, which can still be consistent with gravitino DM for sufficiently small RPV~\cite{Takayama:2000uz, Buchmuller:2007ui}.  Alternatively, the LOSP may dominantly decay to a light hidden sector, which, if thermalized, may not produce significant gravitinos due to its lighter mass scale.  A third possibility is that the LOSP is colored, in which case a late annihilation stage, after the QCD phase transition, can dilute the abundance of $R$-hadrons~\cite{Kang:2006yd, Jacoby:2007nw} before the LOSP decays to gravitinos.  In these cases, a bound on $\tilde{m}$ results from dropping the FO term and is shown on the right of Fig.~\ref{fig:SingleScaleVariations}.  The maximal $\tilde{m}$ occurs at $m_{3/2} = \tilde m$, when Eq.~\eqref{eq:gravDM} gives
\be
\tilde m^2 \, \leq \,  \frac{a}{C_D} \; T_{eq} M_{pl} \,   \lesssim \, (10^3~\mathrm{TeV})^2.
\ee
The numerical value above was obtained for $T_R=\tilde m$.   For larger reheat temperatures the bound is stronger. 

A more drastic possibility is to consider an LSP that violates both assumptions (iv-A,B) entirely,    {\it i.e.} a state that is not the gravitino and yet interacts with the observable sector so weakly that it remains out of equilibrium.  An example is a light, weakly coupled singlino.  In this case, the bound can be completely removed as the singlino couplings can be chosen to be arbitrarily small (removing scattering and freeze-in production) simultaneously with a vanishing mass (thereby removing freeze-out and decay), allowing arbitrarily heavy superpartners.  The key characteristic about the gravitino that leads to our bound is that its mass is inversely related to its coupling to observable states, so that the mass and coupling cannot simultaneously be taken too small.

%%%%%%%%%%%%%%%%%%%%%%%%%%%%%%%%%%%%%%
{\noindent\bf GRAVITY MEDIATION}. 
%%%%%%%%%%%%%%%%%%%%%%%%%%%%%%%%%%%%%%
\begin{figure}[t!]
\centering
\includegraphics[width=0.8\linewidth]{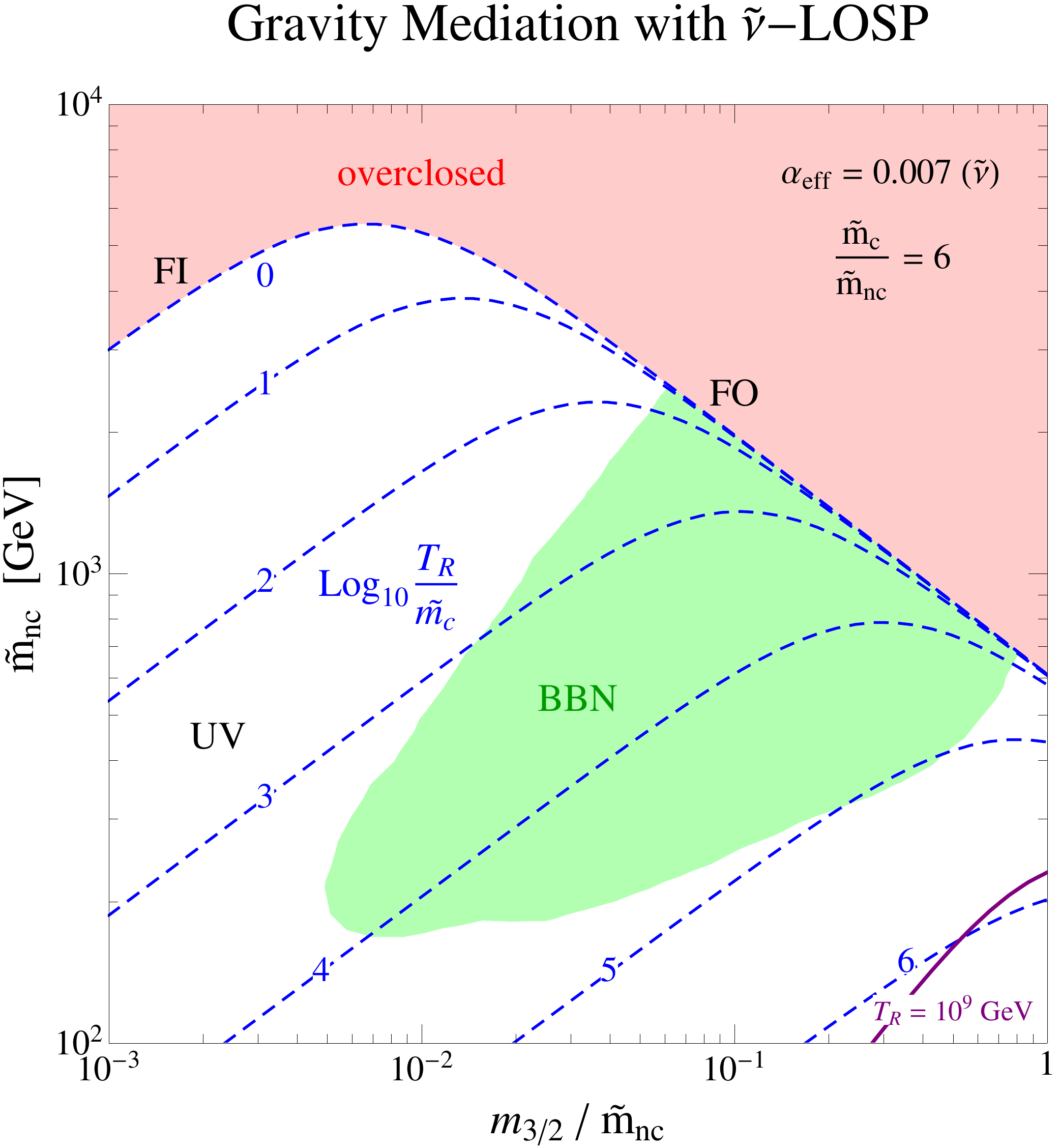}
\caption{\label{fig:sneutrino} 
A bound on $\tilde{m}_{nc}$ for a sneutrino LOSP with gravity-mediated supersymmetry breaking and $\tilde{m_c}/ \tilde{m}_{nc} = 6$ and various $T_R$.  The dashed blue lines show the bound for different values of $T_R$ normalized to $\tilde m_c$.   The purple line at the bottom-right corner shows the bound for $T_R=10^9$ GeV, corresponding to the rough reheat temperature required for successful thermal Leptogenesis. }  
\end{figure}
When mediation of supersymmetry breaking occurs at a very high fundamental scale, $M_*$, of order the scale of gauge coupling unification or higher, then $m_{3/2}/\tilde{m} \sim M_*/M_{Pl} \sim 10^{-3} - 1$.   Thus ``gravity mediation" typically has a gravitino LSP and selects a small region of Fig.~\ref{fig:SingleScale} that is within a few orders of magnitude of the $m_{3/2} = \tilde{m}$ dashed line.  Part of this region, with $M_*$ near $M_{Pl}$, is typically highly constrained by BBN, but smaller values of $M_*$ are of interest and include the largest values of $\tilde{m}$.  

The details of this gravity-mediated region are highly dependent on the LOSP, the superpartner spectrum and $T_R$.  In Fig.~\ref{fig:sneutrino} we show a particular example: a sneutrino LOSP with $\tilde{m}_c/\tilde{m}_{nc} = 6$.    BBN is affected dominantly by rare sneutrino decays with a radiated $Z$ or $W$, so the excluded green region is quite small~\cite{Kawasaki:2008qe, Kanzaki:2006hm}, allowing various possibilities.  One has a light, e.g. 200 GeV, sneutrino,  with $M_*$ near $M_{Pl}$ and a high $T_R \sim 10^9$ GeV, compatible with Leptogenesis~\cite{Davidson:2002qv}.  In this case the colored superpartners may well be in reach of the LHC.  Another possibility  has $M_*$ further from $M_{Pl}$ and a much lower $T_R$ so that the sneutrino mass can be near its upper bound of 5 TeV.

%\begin{acknowledgments}
{\it\bf Acknowledgments:} 
We thank Csaba Csaki, Maxim Perelstein, and Raman Sundrum for helpful conversations.
We also thank the Galileo Galilei Institute for Theoretical Physics for hospitality when this work was initiated.
This work was supported in part by 
the US Department of Energy under 
Contract DE-AC02-05CH11231 and by the National Science Foundation under 
grants PHY-0457315 and PHY-0855653.  J.T.R. is supported by a fellowship from the Miller Institute for Basic Research in Science.  T.V. is supported in part by a grant from the Israel Science Foundation, the US-Israel Binational Science Foundation, and the EU-FP7 Marie Curie, CIG fellowship.

%\end{acknowledgements}
%%%%%%%%%%%%%%%%%%%%%%%%%%%%%%%%%%%%%%%%%%%%%%

 %%%%%%%%%%%%%%%%%%%%%%%%%%%%%%%%%%%%%%%%%%%%%%
%%%%%%%%%%%%%%%%%%%%%%%%%%%%%%%%%%%%%%%%%%%%%%
\end{document}